\documentclass[twocolumn,showpacs,preprintnumbers,amsmath,amssymb]{revtex4}
\usepackage{dcolumn}
\usepackage{bm}
\usepackage{graphicx}%
\usepackage{epsfig}
\begin{document}
\title{A reservoir for inverse power law decoherence of a qubit}
\author{Filippo Giraldi}
\email{giraldi@ukzn.ac.za, filgi@libero.it} \affiliation{Quantum Research Group,
School of Physics and National Institute for Theoretical Physics,
University of KwaZulu-Natal, Durban 4001, South Africa}
\author{Francesco Petruccione}
\email{petruccione@ukzn.ac.za}
\affiliation{Quantum Research Group, School of Physics and National Institute for Theoretical Physics,
University of KwaZulu-Natal, Durban 4001, South Africa}

\begin{abstract}
The exact dynamics of a Jaynes-Cummings model for a qubit interacting with a continuous distribution of bosons, characterized by a special form of the spectral density, is evaluated analytically. The special reservoir is designed to induce anomalous decoherence, resulting in an inverse power law relaxation, of power $3/2$, over an evaluated long time scale. If compared to the exponential-like relaxation obtained from the original Jaynes-Cummings model for Lorentzian-type spectral density functions, decoherence is strongly suppressed. The special reservoir
exhibits a band edge frequency coinciding with the qubit transition frequency. Known theoretical models of photonic band gap media suitable for the realization of the designed reservoir are proposed.
\end{abstract}

\maketitle
\section{Introduction}

The dynamics of a two-level system (TLS) coupled to an  environment is of great interest \cite{WW,legett}, given the many applications in quantum optics, quantum information, atom-cavity interactions, molecular dynamics, spectroscopy, and solid state physics. Within the theory of open quantum systems
  \cite{BP} the environment is often represented as a reservoir of bosons \cite{weiss}, characterized by the corresponding spectral density function. Several studies have been performed to investigate the dependance of decoherence on the explicit form of the spectral density function. According to the system of relevance, sub-ohmic, ohmic, super-ohmic and Lorentzian forms of the spectral density have been investigated \cite{BP, weiss}, by adopting the celebrated Jaynes-Cumming model \cite{jc}.

     Interesting results emerge from the technique of the resolvent operator \cite{ctbook} applied to a two-level atom (TLA) interacting, in rotating wave approximation, with a Lorentzian \cite{szbook} distribution of field modes. The coupling constants are assumed to vary slowly for frequency changes. In this way, oscillating behaviors enveloped in exponential decays emerge in the exact dynamics. For a detailed report we refer to \cite{lambropoulos}.

     Another model mimicking the spontaneous decay of a TLA in a structured reservoir, has been introduced by Garraway \cite{garraway,garraway2} and solved exactly for Lorentzian type distributions and a special non-Lorentzian one with two poles in the lower half plane. The assumption that the frequency range is $\left(-\infty,+\infty\right)$
  makes it possible to characterize the exact dynamics through the poles of the spectral density in the lower half plane and the corresponding pseudomodes \cite{garrawayknight}. The resulting dynamics is described by oscillations enveloped in exponential relaxations.

  Recently, the model adopted in Ref. \cite{garraway} has been resumed to compare the exact master equation in time-convolutionless form with the Nakajima-Zwanzig master equation, where the perturbation expansion of the corresponding memory kernel is performed \cite{VB}.

In line with the attempt to delay the destructive effects of the environment on the qubit, i.e., decoherence, we consider the model adopted in Ref. \cite{garraway} and study a reservoir that is piecewise similar to those usually adopted, e.g.,  sub-ohmic and  Lorentzian ones. We show the relevant changes in the time evolution induced by the specially designed reservoir and how the decoherence process results to be strongly suppressed. We notice that the analytical calculations concerning the exact dynamics, are performed for a positive range of modes frequencies.

\section{The model}

The model considered was originally introduced by Garraway \cite{garraway} in order to describe the exact dynamics of an atom interacting
with a cavity field, approximated by a continuous distribution of modes,
described by a Lorentzian, a linear combination of two Lorentzian and a non-Lorentzian distribution. Briefly, the
Hamiltonian of the whole system reads $H=H_S+H_E+H_I$, where the system Hamiltonian $H_{S}$, the bath Hamiltonian $H_{E}$ and the interaction Hamiltonian are given by the following forms:
\begin{equation}
H_S=\omega_0\, \sigma_+\sigma_-,
\hspace{2em}H_E=\sum_{k=1}^{\infty} \omega_k \,a^{\dagger}_k a_k,\nonumber
\end{equation}
\begin{equation}
H_I=
\sum_{k=1}^{\infty} \left(g_k \,\sigma_+\otimes a_k+g_k^{\ast}\,
\sigma_-\otimes a^{\dagger}\right),\nonumber
\end{equation}
being $\hbar=1$ in the system of units adopted.
  In the usual notation $\sigma_+$ and $\sigma_-$ are the rising and lowering operators, respectively, acting on the Hilbert space of the qubit,
   while $a_k^{\dagger}$ and $a_k$ are the creation and annihilation operators,
   respectively, acting on the Hilbert space of the $k$-th boson, satisfying the boson commutation rule,
   $\left[a_k,a_{k^{\prime}}^{\dagger}\right]=\delta_{k,k^{\prime}}$, for every non-vanishing natural value of
   $k$ and $k^{\prime}$. The qubit transition frequency is $\omega_0$, while the constants $g_k$ represent the coupling between the $k$-th field mode and the qubit transition.
   As follows we refer to a TLS interacting with a cavity supplying a reservoir of field modes studied by Garraway \cite{garraway} and adopted in Ref. \cite{VB}.

   Starting from an initial generic unentangled condition where the cavity is in the vacuum state $|0\rangle_E$,
   \begin{equation}
|\Psi(0)\rangle=\left(c_0 |0\rangle+ c_1(0)|1\rangle\right) \otimes |0\rangle_E, \nonumber
\end{equation}
   the exact time evolution is
   described by the form
   \begin{equation}
|\Psi(t)\rangle=c_0 |0\rangle \otimes |0\rangle_E+c_1(t)|1\rangle\otimes|0\rangle_E+\sum_{k=1}^{\infty}b_k(t)|0\rangle \otimes |k\rangle_E, \nonumber
\end{equation}
 where $|k\rangle_E=a^{\dagger}_k |0\rangle_E$ for every $k=0,1,2,\ldots$.
 The dynamics is easily studied in the interaction picture,
   \begin{equation}
   \begin{split}
      &|\Psi(t)\rangle_I=  e^{\imath \left(H_S+H_E\right)t}|\Psi(t)\rangle \nonumber \\
&=c_0 |0\rangle \otimes |0\rangle_E+C_1(t)|1\rangle\otimes|0\rangle_E+\sum_{k=1}^{\infty}B_k(t)|0\rangle \otimes |k\rangle_E, \nonumber \\ & C_1(t)=e^{\imath \omega_0 t}c_1(t),\hspace{1em} B_k(t)=e^{\imath \omega_k t}b_k(t), \hspace{1em} k=1,2,\ldots,
\end{split}
\end{equation}
where $\imath$ is the imaginary unit. The amplitude $C_1(t)$ is driven by the convoluted structure equation
\begin{equation}
\dot{C}_1(t)=-\left(f\ast C_1\right)(t),
\label{cMEQ&CorrReservoir}
\end{equation}
 where  $f$ is the two-point correlation function of the reservoir of field modes,
 \begin{equation}
f\left(t-t^{\prime}\right)=\sum_{k=1}^{\infty}
\left|g_k\right|^2 e^{-\imath \left(\omega_k-\omega_0\right)\left(t-t^{\prime}\right)}. \nonumber
\end{equation}
  For a continuous distribution of modes described by $\eta\left(\omega\right)$, the correlation function is expressed through the spectral density function $J\left(\omega\right)$,
 \begin{equation}
f\left(\tau\right)=
\int_0^{\infty}J\left(\omega\right) e^{-\imath\left(\omega-\omega_0\right)\tau }
d \omega, \nonumber
\end{equation}
where $J\left(\omega\right)=\eta\left(\omega\right) \left|g\left(\omega\right)\right|^2$ and $g\left(\omega\right)$ is the frequency dependent coupling constant.

The reduced density matrix, obtained by tracing over the degrees of freedom  of the Hilbert space of the bosons, reads
\begin{equation}
\rho_{1,1}(t)=1-\rho_{0,0}(t)=\rho_{1,1}(0)\,\left|G(t)\right|^2,
\label{rhot11}
\end{equation}
\begin{equation}
\rho_{1,0}(t)=\rho_{0,1}^{\ast}(t)=\rho_{1,0}(0)\,e^{-\imath \omega_0 t  }G(t).\label{rhot10}
\end{equation}
The term $G(t)$ fulfills the convolution equation:
\begin{equation}
\dot{G}(t)=-\left(f\ast G\right)(t),\hspace{1em}G(0)=1.
\label{G}
\end{equation}

\section{the Exact dynamics}

We study the exact dynamics of the reduced density matrix of the qubit as given by Eq. (\ref{G}) for a reservoir described by the following spectral density function:
\begin{equation}
J\left(\omega\right)=
 \frac{2 A \left(\omega-\omega_0\right)^{1/2}
 \Theta\left(\left(\omega-\omega_0\right)/\omega_0\right)}
 {a^2+\left(\omega-\omega_0\right)^2 },
\label{JE}
\end{equation}
where $\Theta$ is the Heaviside step function, while $A$ and $a$ are arbitrary positive constants. Note that $J\left(\omega\right)$ exhibits a sub-ohmic behavior at low frequencies and has an absolute maximum, $3^{3/4} A /\left(2 \,a^{3/2}\right)$, at the frequency $\omega_M=\omega_0+a/\sqrt{3}$. In contrast, the high frequency behavior of $J\left(\omega\right)$ is an inverse power law like the usual Lorentzian one, though with a different power.
The \emph{exact} solution of Eq. (\ref{G}) corresponding to the spectral density (\ref{JE}), can be shown to be
  \begin{equation}
 G(t)=\frac{1}{ \sqrt{\pi}}\sum_{l=1}^4
  R\left(z_l\right)\,
 z_l\,e^{ z_l^2 t}\,\Gamma\left(1/2,z_l^2 t\right),\label{Gt}
\end{equation}
  where $R(z)$ is a rational function,
  \begin{equation}
   R(z)=\frac{\left(1-\imath\right)
   \left(a^{1/2}+z\right)\left(\imath \,a^{1/2}+ z\right)}
   { 2 z \left(\left(1+\imath\right)a+3 a^{1/2} z +2 \left(1-\imath\right)z^2\right) },
   \label{R}
   \end{equation}
  while the complex numbers $z_1,z_2,z_3$ and $z_4$, are the roots, \emph{distinct} for every positive value of both $A$ and $a$, of the polynomial $Q(z)$, given by the following form:
   \begin{equation}
    Q(z)=\pi \sqrt{2/a} \,A +\imath \,a \,z^2+ \left(1+\imath\right)a^{1/2} z^{3}+z^4. \label{Q}
   \end{equation}
  For the sake of shortness, we do not report the analytical expressions of the roots, while the proof that the expression (\ref{Gt}) is solution of Eq. (\ref{G}) is summarized in Appendix \ref{A}.
    The asymptotic expansion of the Incomplete Gamma functions \cite{em2} in Eq. (\ref{Gt}), identifies a \emph{time scale} $\tau$ and a \emph{decoherence factor} $\mathcal{D}$, depending on both the parameters $A$ and $a$,
  \begin{equation}
 \tau=\max \left\{\left|z_l\right|^{-2},\,l=1,2,3,4\right\},
 \label{tauE}
 \end{equation}
  \begin{equation}
 \mathcal{D}=\frac{1}{2\sqrt{\pi} }
  \sum_{l=1}^4  R\left(z_l\right)\,
 z_l^{-2}, \label{DE}
 \end{equation}
 such that, for times $t\gg\tau$,
 a power law behavior emerges, described by the following asymptotic form:
\begin{equation}
 G(t)\sim-\mathcal{D} \,t^{-3/2}
  , \hspace{2em} t\to+\infty. \label{Gasympt}
 \end{equation}

 We are finally equipped to give the \emph{exact} time evolution of the reduced density matrix through Eq. (\ref{rhot11}) and Eq. (\ref{rhot10}),
 \begin{equation}
\rho_{1,1}(t)=
\frac{\rho_{1,1}(0)}{ \pi}
\Bigg| \sum_{l=1}^4
R\left(z_l\right)\, z_l\,e^{ z_l^2 t}\,\Gamma\left(1/2,z_l^2 t\right)\Bigg|^2,
\label{rho11}
\end{equation}
\begin{equation}
 \rho_{1,0}(t)  =
\frac{\rho_{1,0}(0)}{ \sqrt{\pi}}\,e^{-\imath \omega_0 t}\,\sum_{l=1}^4
R\left(z_l\right)\, z_l\,e^{ z_l^2 t}\,\Gamma\left(1/2,z_l^2 t\right),
\label{rho10}
\end{equation}
 obviously, $\rho_{0,0}(t)=1-\rho_{1,1}(t)$ and $\rho_{0,1}^{\ast}(t)=\rho_{1,0}(t)$.

 Over long time scales, $t \gg \tau$, the dynamics is described by the asymptotic forms obtained through Eq. (\ref{Gasympt}),
  \begin{equation}
\rho_{1,1}(t)\sim
\rho_{1,1}(0)
\left| \mathcal{D}\right|^2t^{-3},\hspace{2em}t\to+\infty,\label{rho11asympt}
\end{equation}
\begin{equation}
\rho_{1,0}(t)\sim-
\rho_{1,0}(0)\,e^{-\imath \omega_0 t}\,\mathcal{D}\,t^{-3/2}
,\hspace{2em} t \to+\infty.
\label{rho10asympt}
\end{equation}
Ultimately, the qubit  collapses into the ground state.

The dynamics of the qubit, driven by the special reservoir described by the spectral density function $J\left(\omega\right)$, results in a linear combination of Incomplete Gamma functions. In contrast to the \textit{exponential}-like relaxation occurring for a Lorentzian spectral density function, in the present case the  relaxation is \textit{Eulerian}.
The time scale $\tau$ suggests the times for inverse power law relaxations with powers either $3/2$ or $3$, for either the coherence term or the populations, respectively.

\section{Exponential vs Eulerian relaxation }

It is now useful to compare the above results to the dynamics of qubit in a Garraway model \cite{VB,BP,garraway}, where the reservoir of field modes is described by a Lorentzian spectral density function
\begin{equation}
J_L\left(\omega \right)= \frac{1}{2 \pi}\frac{\gamma \lambda^2}{\left(\omega-\omega_0\right)^2+\lambda^2}.
\end{equation}
The corresponding bath correlation function is
\begin{equation}
f_L(t)= \frac{\gamma \lambda }{2} \, e^{-\lambda \left|t\right|},
\label{jLfL}
\end{equation}
the positive constant $\lambda $ defines the spectral width of the coupling and is the inverse of the reservoir correlation time $\tau_B$, while the positive constant $\gamma$ is the inverse of $\tau_R$, the time scale on which the system changes. For details we refer to \cite{BP}.
The time evolution of the reduced  density matrix of the qubit reads
\begin{equation}
 \rho_{1,1}(t)=\rho_{1,1}(0)\,
\left|G_L(t)\right|^2,  \label{rho11L}
\end{equation}
\begin{equation}
\rho_{1,0}(t)=\rho_{1,0}(0)\,e^{-\imath \omega t  } G_L(t).\label{rho10L}
\end{equation}
For $\lambda> 2 \gamma$, we have
\begin{equation}
 G_L(t)=e^{-\lambda t/2} \left(\cosh \left(\frac{   d }{2}\,t\right)
+\frac{\lambda}{ d}\sinh \left(\frac{  d }{2}\,t\right)\right), \nonumber
\end{equation}
being $d=\sqrt{\lambda^2- 2 \gamma \lambda}$, while, for $\lambda> 2 \gamma$,
\begin{equation}
G_L(t)=e^{-\lambda t/2} \left(\cos \left(\frac{  \hat{d} }{2}\,t\right)
+\frac{\lambda}{ \hat{d}}\sin \left(\frac{  \hat{d} }{2}\,t\right)\right), \nonumber
\end{equation}
being $ \hat{d}=\sqrt{ 2 \gamma \lambda-\lambda^2}$, with discrete zeros at times $t_n=2/\hat{d}\left(n \pi-\arctan\left(\hat{d}/\lambda\right)\right)$, for every $n=1,2,\ldots$.

We now compare the exponential-like decoherence process, given by a reservoir described by the Lorentzian spectral density function $J_L\left(\omega\right)$, to the Eulerian relaxation, resulting in an inverse power law over long time scales, corresponding to $J\left(\omega\right)$, given by Eq. (\ref{JE}). A long time scale inverse power law relaxation is slower than an exponential one, i.e., the decoherence process is strongly suppressed by adopting reservoirs described by $J(\omega)$, instead of $J_L(\omega)$. The qubit has "longer life". Numerical computations (Mathematica) for particular values of the parameters are performed. For $A=0.8 \,a^{5/2}$, $\omega_0=a/2$, the following estimates emerge: $
  z_1 \,a^{-1/2}\simeq-1.282+\imath \,0.716$,
  $z_2\,a^{-1/2}\simeq-1.150-\imath\, 1.150$, $z_3\,a^{-1/2}\simeq 0.716-\imath \,1.282$, $z_4\,a^{-1/2}\simeq 0.717+\imath\, 0.717$, $ \tau\simeq 0.974/a$, $\left|\mathcal{D}\right|\simeq 0.112 \,a^{-3/2}$ and the dynamics of the qubit can be evaluated starting, for example, from the initial conditions $\rho_{1,0}(0)=1/5$ and $\rho_{1,1}(0)=1/2$.
\begin{figure}[t]
\centering
\includegraphics[height=4.25 cm, width=7.25 cm]{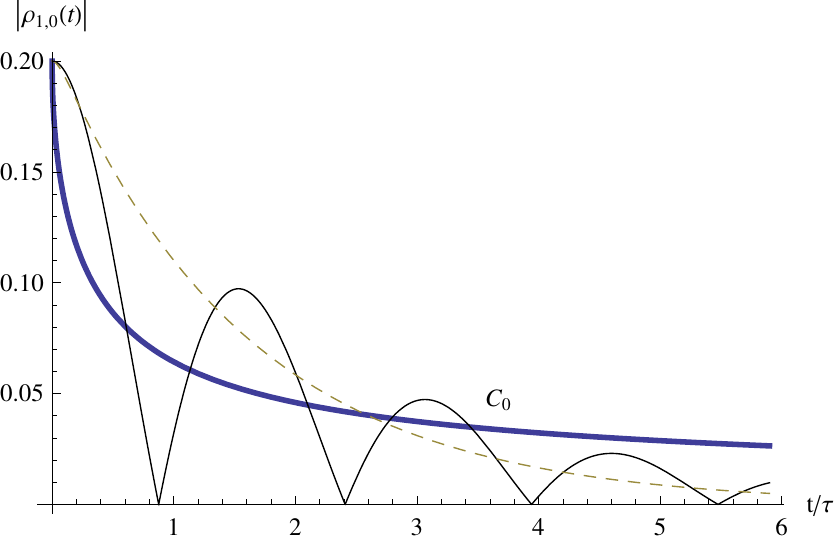}
\vspace{0.2cm}
\caption{The coherent term $\left|\rho_{1,0}(t)\right|$ of the reduced density matrix of a qubit interacting with a reservoir, described by either $J_L\left(\omega\right)$, both in strong (thin solid line) and weak coupling regime (dashed line), or $J\left(\omega\right)$ (thick solid line) spectral density function, given by Eq. (\ref{rho10}) and Eq. (\ref{rho10L}), respectively, for $0\leq t/\tau\leq 5.9$.
The values of the parameters are $A=0.8\,a^{5/2}$, $\omega_0=a/2$, $ \tau\simeq 0.974 /a$, $\tau_B=1/a$ and $\tau_R=1/(10 \,a)$ in strong coupling regime, $ \tau_B=1/(20\, a)$ and $\tau_R=10/(13\,a)$ in weak coupling regime, the initial conditions read $\rho_{1,1}(0)=1/2$, $\rho_{1,0}(0)=1/5$, and $C_0$ labels the cross point of the largest time coordinate between the thin and the thick solid line.}
\label{R10initial}
\end{figure}
\begin{figure}[t]
\centering
\includegraphics[height=4.25 cm, width=7.25 cm]{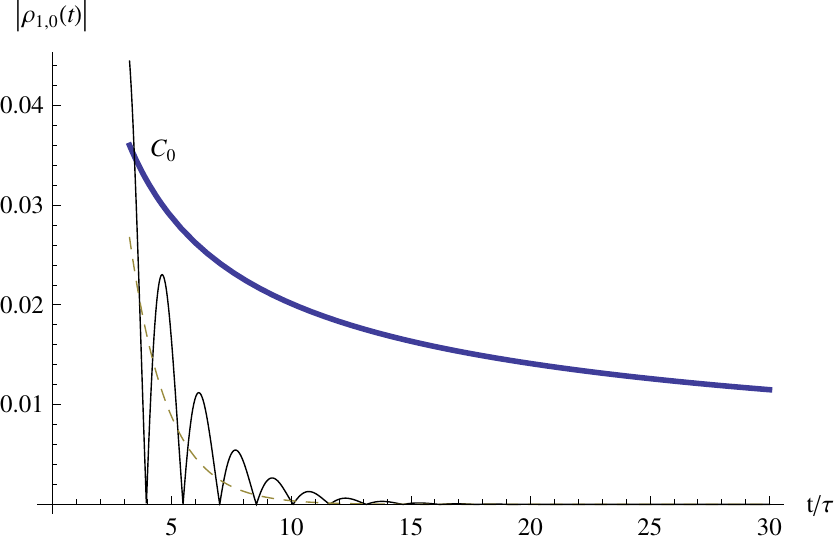}
\vspace{0.2cm}
\caption{ The relaxation of coherent term $\left|\rho_{1,0}(t)\right|$ over long time scales, of the reduced density matrix of a qubit, interacting with a reservoir, described by either $J_L\left(\omega\right)$, both in strong (thin solid line) and weak coupling regime (dashed line), or $J\left(\omega\right)$ (thick solid line) spectral density function, given by Eq. (\ref{rho10asympt}) and Eq. (\ref{rho10L}), respectively, for $3.2\leq t/\tau \leq30$.
The values of the parameters are $A=0.8\,a^{5/2}$, $\omega_0=a/2$, $ \tau\simeq 0.974/a $, $\tau_B=1/a$ and $\tau_R=1/(10 \,a)$ in strong coupling regime, $\tau_B=1/(20\,a)$ and $\tau_R=10/(13\,a)$ in weak coupling regime, the magnitude of the decoherence factor reads $\left|\mathcal{D}\right|\simeq 0.112\, a^{-3/2}$, the initial conditions are $\rho_{1,1}(0)=1/2$, $\rho_{1,0}(0)=1/5$, and $C_0$ labels the cross point of the largest time coordinate between the thin and the thick solid line.}
\label{R10asympt}
\end{figure}

 For the damped Jaynes-Cumming model the absolute value of the coherence term shows an oscillating behavior enveloped in an exponential decay in strong coupling regime, $\gamma>\lambda/2$, and a linear combination of exponential decays in weak coupling regime, $\gamma<\lambda/2$, while, the Eulerian dynamics exhibits a power law decay proportional to $t^{-3/2}$ over long time scales, $t \gg \tau$.
 A detailed behavior is shown in Fig. ~\ref{R10initial} for a short time scale, $t \lesssim \tau $, and in Fig. ~\ref{R10asympt} for long time scales, $t\gg \tau$, respectively.

 In agreement with the exact analytical results obtained, the reservoir described by $J\left(\omega\right)$ gives an inverse power law relaxation, with power $3/2$, over long time scales, $t\gg\tau$, given by Eq. (\ref{tauE}). If compared to the exponential-like decay corresponding to a reservoir described by $J\left(\omega\right)$, the decoherence process is slowed.

\section{Conclusions}

The exact dynamics of a TLS in the Jaynes-Cummings model, has been determined analytically by Garraway for Lorentzian type and a non-Lorentzian spectral density functions of the reservoir of bosons, in each case decoherence results in an exponential-like relaxation. In line with the attempt to slow down the decoherence process, we design a reservoir piecewise similar to those usually adopted, i.e., a $1/2$ power law behavior at low frequencies, similar to the sub-ohmic case, and a $3/2$ inverse power law behavior at high frequencies, similar to the Lorentzian case, though with a different power. The special reservoir also exhibits a photonic band gap (PBG) edge \cite{jpc,jpbgm} coinciding with the qubit transition frequency.

The \emph{analytical} description of the \emph{exact} time evolution of the qubit is a combination of Eulerian functions of time and decoherence results in a \emph{$3/2$ inverse power law relaxation} over an evaluated \emph{long time scale}. Ultimately, the system collapses into the ground state. If compared to the relaxation process corresponding to a Lorentzian type spectral density function, the process of decoherence is \emph{strongly suppressed}.

An environment implementing the specially designed reservoir of modes can in principle be realized with PBG media \cite{y87,j87}. An anisotropic model providing a PBG close to the 3D photonic crystals \cite{jpc}, is discussed in Refs. \cite{jw0} and \cite{jw1}. The corresponding density of modes reads \[\eta\left(\omega\right)\propto \sqrt{\omega-\omega_e} \,\, \Theta\left(\left(\omega-\omega_e\right)/\omega_e\right),\] where $\omega_e$ is the band edge frequency.
 If the qubit transition frequency coincides with the edge of the PBG, the low frequency behavior of the specially designed reservoir is recovered by assuming that the couplings vary slowly at low frequencies, $g\left(\omega\right)\simeq g\left(\omega_0\right)$ for $\omega \gtrsim \omega_0$.
Physically, the modes relevant for the dynamics are the resonant ones, which means that the time evolution mostly depends on the frequency behavior near the transition frequency of the system of interest \cite{GZ}. If a qubit can be placed in such a material
 and interact with such a reservoir (in rotating wave approximation), the described decoherence process can in principle appear. Of course, the accuracy of the model  depends upon the high frequency behavior of the spectral density.

Another theoretical model providing a structured PBG is the $N$-period one dimensional lattice discussed in Ref. \cite{bendickson}. The density of frequency modes can reproduce a band gap by properly arranging the periodic sequence of unit lattice cells.  The density of modes is evaluated analytically as a function of the complex transmission coefficients of each unit cell.

 Another potential realization of structured PBG are tunable 1D PBG microcavities \cite{PBGMC1,PBGMC2}. Their fabrication is achieved through advanced diffractive grating and photonic crystals technologies.

The action of such structured PBG environments on a qubit could be a way of delaying the decoherence process with fundamental applications to Quantum Information Processing Technologies. In future we plan to search for specially engineered reservoirs inducing an even slower relaxation of qubit decoherence.

\appendix
\section{the time evolution in details}\label{A}

A detailed analysis of the equations driving the exact dynamics, is performed and the analytical solution of Eq. (\ref{G}) is evaluated in case the reservoir is described by the spectral density (\ref{JE}).

The following class of non-negative, non-divergent and summable spectral density functions:
\begin{equation}
  \int_0^{\infty}J\left(\omega\right)
d \omega<\infty,\hspace{1em} J\left(\omega\right)=\Theta\left(\left(\omega-\omega_0\right)/\omega_0\right)
\Lambda\left(\omega-\omega_0\right),\nonumber
\end{equation}
gives the Laplace transform
\begin{equation}
\tilde{G}(u)=\left[u-\imath\,\mathcal{S}
\left(\Lambda\right)\left(-\imath u\right)\right]^{-1},
\label{Gu}
\end{equation}
as a function of the Stieltjes transform $\mathcal{S}
\left(\Lambda\right)(u)$, holding true in case $\Re \left\{u\right\}>0$ and
$\left|\arg\left\{-\imath u\right\}\right|<\pi$.
The fundamental uniform convergence of integrals involved is guaranteed by the constraints:
$\int_0^{\infty} \Lambda\left(\omega\right)d \omega<\infty$ and
$\Re\left\{u\right\}>0$.
We also notice that we adopt a positive range of continuous frequencies.
The choice of the spectral density function (\ref{JE}) leads to the following Laplace transform:
\begin{equation}
 \tilde{G}(u)=\frac{a^{1/2} \left(\imath a^{1/2}+u^{1/2}\right)\left(a^{1/2}+u^{1/2}\right)
  }{\left(\pi \sqrt{2} A +\imath a^{3/2}u+ \left(1+\imath\right)a u^{3/2}
 +a^{1/2}u^2 \right)}.\nonumber
 \end{equation}
 We are finally equipped to find $G(t)$, solution of Eq. (\ref{G}). The analysis of the discriminant shows that the roots of the polynomial $Q(z)$, given by Eq. (\ref{Q}), are \emph{distinct} for every value $A>0$ and $a > 0$, thus, the Inverse Laplace Transform \cite{NM} of $\tilde{G}(u)$,
 \begin{equation}
 G(t)=\frac{t^{-3/2}}{2 \sqrt{\pi}} \sum_{l=1}^4
 R\left(z_l\right)\int_0^{\infty}
  \tau e^{z_l \tau-\tau^2/\left(4 t\right)} d \tau\nonumber,\quad t>0,
  \end{equation}
 leads to Eq. (\ref{Gt}). Also, the relations
    \begin{equation}
 \sum_{l=1}^4 R \left(z_l\right)=0,\hspace{2em}\lim_{t\to0^+} G(t)=\sum_{l=1}^4  R\left(z_l\right) \, z_l=1, \nonumber
  \end{equation}
  obtained through the Initial value theorem, allow the continuation $G(0)=1$. Thus, Eq. (\ref{Gt}) is the \emph{exact} solution of Eq. (\ref{G}), where $R(z)$ is given by Eq. (\ref{R}).

  Finally, the asymptotic expansion of the incomplete Gamma function gives Eq. (\ref{Gasympt}), describing the behavior of $G(t)$ over long time scales, $t\gg\tau$, defined by Eq. (\ref{tauE}), driving the dynamics over long time scale, as well.

 \begin{acknowledgments}
 This work is based upon research supported by the South African Research Chairs Initiative of
the Department of Science and Technology and National Research Foundation.
\end{acknowledgments}

\end{document}